\documentclass[prl,twocolumn,floatfix,superscriptaddress]{revtex4-2}

\usepackage{booktabs}
\usepackage{amsmath,amssymb,amsthm,bm}
\usepackage{bbm}
\usepackage{braket}   

\usepackage{graphicx}
\usepackage{tikz}

\usepackage{multirow}
\usepackage{enumitem}
\usepackage{xcolor,subfigure,makecell}

\usepackage[colorlinks]{hyperref}
\hypersetup{
  bookmarksnumbered,
  pdfstartview={FitH},
  citecolor=blue,
  linkcolor=red,
  urlcolor=black,
  pdfpagemode=UseOutlines
}
\usepackage{cleveref}

\usepackage{xparse}

\usepackage[newcommands]{ragged2e}

\theoremstyle{definition}

\allowdisplaybreaks

\newcommand{\ceil}[1]{\lceil {#1} \rceil}

\makeatletter
\newenvironment{proof*}[1][\proofname]{\par
  \pushQED{\qed}%
  \normalfont \partopsep=\z@skip \topsep=\z@skip
  \trivlist
  \item[\hskip\labelsep\itshape #1\@addpunct{.}]\ignorespaces
}{%
  \popQED\endtrivlist\@endpefalse
}
\makeatother

\ExplSyntaxOn
\NewDocumentCommand{\mref}{m}{\quinn_mref:n {#1}}
\seq_new:N \l_quinn_mref_seq
\cs_new:Npn \quinn_mref:n #1
 {
  \seq_set_split:Nnn \l_quinn_mref_seq { , } { #1 }
  \seq_pop_right:NN \l_quinn_mref_seq \l_tmpa_tl
  ( 
  \seq_map_inline:Nn \l_quinn_mref_seq { \ref{##1},\nobreakspace }
  \exp_args:NV \ref \l_tmpa_tl
  ) 
 }
\ExplSyntaxOff

\begin{document}
\title{Quantifying the Operational Cost of Multipartite Entanglement}
\author{Francois Payn}
\email{francois.payn00@gmail.com}
\affiliation{DISAT, Politecnico di Torino,  Torino 10129, Italy}
\author{Michele Minervini}
\affiliation{School of Electrical and Computer Engineering,
Cornell University, Ithaca, New York 14850, USA
}
\author{Davide Girolami}
\email{davegirolami@gmail.com}
\affiliation{DISAT, Politecnico di Torino,  Torino 10129, Italy}
\date{\today}

\begin{abstract}
Multipartite entanglement determines the strength and range of interactions in many-body quantum systems. Yet,   
it is hard to evaluate it, due to the complex structures of quantum states. Here, we introduce a generic method  to quantify the $k\leq N$-partite entanglement of an $N$-particle system, by maximizing an arbitrary bipartite entanglement measure within subsystems of size up to $k$. The resulting classification of multipartite states captures their experimental cost: creating a $k$-partite entangled state requires at least $k-1$ quantum communication channels. Further, we analytically calculate the newly defined $k$-partite entanglement of formation, which generalizes an important bipartite entanglement measure, in several classes of states, including the $W$ states of any dimension.
\end{abstract}

\maketitle

{\it Introduction --}
Entanglement is  the most important property of interacting quantum systems and the key resource for many sought-after quantum technologies \cite{rev,Jozsa-Linden,entapp1,comm,vidal,book}. In particular,   large-scale quantum computers and  quantum communication networks rely on {\it multipartite} entanglement, establishing quantum communication among several  particles or laboratories. \\
Yet, its characterization is hard.  A single notion of quantumness cannot fully describe the rich correlation patterns  of quantum systems, whose degrees of freedom  increase exponentially with their size.  \\

\begin{figure}
\includegraphics[width=.5\textwidth,height=5cm]{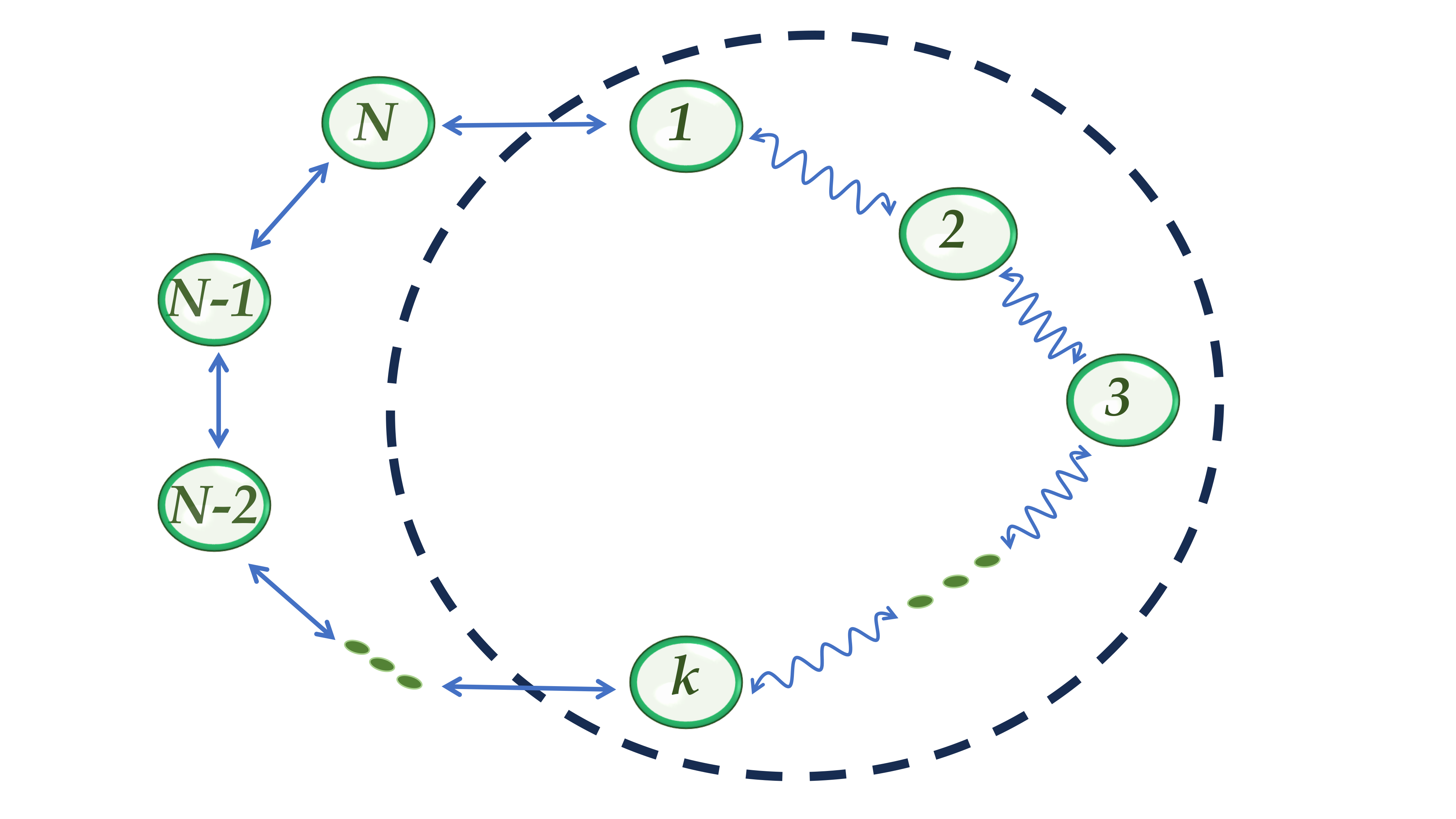}
\caption{We introduce an operational notion of multipartite entanglement: it takes at least $k-1$ bipartite  quantum communication channels (wavy lines)  to create $k$-partite entanglement in an $N$-particle system. We assume that LOCCs (continuous lines) are a free resource.}
\label{fig1}
\vspace{-10pt}
\end{figure}

Here, we introduce a definition of  multipartite entanglement that answers the  important open question: {\it how difficult is it to create a certain entangled state?}  
We  build measures of 
 $k\leq N$-partite entanglement by maximizing the sum of bipartite entanglement terms in subsystems of at most $k$ particles. The resulting  ``total'' entanglement, i.e., the sum of  $k$-partite entanglement of any degree $k$, meets the canonical properties of entanglement monotones. \\
Then, we prove that this notion of multipartite entanglement relates to the experimental cost of quantum states: creating  $k$-partite entanglement values requires  quantum communication between at least $k-1$ particle pairs (FIG.~\ref{fig1}). Therefore, it generalizes the definition of  bipartite entanglement   as information that cannot be exchanged between two particles  via  local operations and classical communication (LOCC) \cite{locc1,locc3}. \\
 The induced classification  of multipartite entangled states is  different from the ones  based on  ``separability'' \cite{sep1,sep2,sep3,sep4,sep5,sep6}, counting the uncorrelated clusters of entangled particles in a system, and ``producibility'' \cite{prod,szalay,szalay0,prod2}, which evaluates the size of those  clusters. Indeed, preparing states in the same separability/producibility class can require very different experimental resources, while the same number of entangling operations can create different states in terms of separability/producibility. \\
 
 Moreover, we  calculate the newly defined  {\it $k$-partite entanglement of formation} in relevant classes of states. Notably, it is analytically computable for $W$ states of any dimension, dissecting their complex entanglement structure \cite{sep1,sep2,rev}. This is surprising, as entanglement quantification usually involves nuanced strategies tailored to the system under scrutiny \cite{entmeasures13,entmeasures1,entmeasures2,entmeasures3,entmeasures4,entmeasures5,entmeasures6,entmeasures7,entmeasures8,entmeasures9,entmeasures10,entmeasures11,entmeasures12}.\\

{\it Defining and Quantifying Multipartite Entanglement --}
Consider a finite dimensional $N$-partite quantum system, $\mathcal{X}_N=\bigl\{ \mathcal{X}_{[1]}\mathcal{X}_{[2]}\dots\mathcal{X}_{[N]} \bigr\}$, in which  $\mathcal{X}_{[i]}$ is the single particle $i$, and $\mathcal{X}_k=\bigl\{ \mathcal{X}_{[1]}, \mathcal{X}_{[2]}, \dots, \mathcal{X}_{[k]} \bigr\}$  is a cluster of $k\leq N$ particles. We call $\rho_{[i]}$ the state of  $\mathcal{X}_{[i]}$, and   $\rho_{k}$ is the state of the cluster $\mathcal{X}_{k}$. 
 The  state of the global system is $\rho_N$.\\
A two-particle state $\rho_2$  is entangled if it cannot be prepared  by LOCCs between $\mathcal{X}_{[1]}$ and $\mathcal{X}_{[2]}$ from an arbitrary initial separable state $\sum_i p_i\rho_{[1],i}\otimes \rho_{[2],i}$.  The exact mathematical characterization of LOCCs is challenging \cite{locc3}, but we know they are a subset of the separable operations $\Phi_{[1][2]}(\rho_{2})=\sum_i K_{[1][2],i}\rho_2 K_{[1][2],i}^\dagger$, in which $K_{[1][2],i}=K_{[1],i}\otimes K_{[2],i}$ are Kraus operators, $\sum_{i}K_{[1][2],i}^\dagger K_{[1][2],i}=I$. \\ 
 Building on such operational definition of bipartite entanglement, we define multipartite entanglement in $\rho_N$ in terms of its difference from states that can be created by LOCCs between different particle pairs from separable states. Specifically, we calculate entanglement {\it up to degree $k$}  by evaluating bipartite entanglement in $N-1$ system clusters of up to $k$ particles, which we identify by   the following iterating procedure. \\
 
First, we select a cluster $\mathcal{X}^1_k =\left\{\mathcal{X}^1_{[1]}\mathcal{X}^1_{[2]}...\mathcal{X}^1_{[k]}\right\}$ of $k\leq N$ particles in $\mathcal{X}_N$. The entanglement between $\mathcal{X}^1_{[1]}$ and the rest of the  cluster is  $E[\rho_N]\left(\mathcal{X}^1_{[1]}:\mathcal{X}^1_{[2]}\mathcal{X}^1_{[3]}... \mathcal{X}^1_{[k]}\right)$, where $E$ is an arbitrary bipartite entanglement quantifier.
The quantity is positive if and only if  the cluster state $\rho^1_{k}$ cannot be created from   $\sum_i p_i\rho^1_{1,i}\otimes \rho^1_{k-1,i}$ by LOCCs between $\mathcal{X}^1_{[1]}$ and the other particles in $\mathcal{X}^1_{k}$.\\
Then,  we choose a second cluster $ \mathcal{X}^2_{k}$ of $k$-particles  that does {\it not} include $\mathcal{X}^1_{[1]}$. Note that in general $\mathcal{X}^1_{[i]}\neq \mathcal{X}^2_{[i]}$ and different particles may be included in the two clusters. The same argument of the first step applies: the condition $E[\rho_N]\left(\mathcal{X}^2_{[1]}:\mathcal{X}^2_{[2]}\mathcal{X}^2_{[3]}... \mathcal{X}^2_{[k]}\right) > 0$ signals that one cannot prepare  the cluster state $\rho^2_k$  by LOCCs between $\mathcal{X}^2_{[1]}$ and the rest of the cluster from $\sum_i p_i\rho^2_{1,i}\otimes \rho^2_{k-1,i}$. \\
Iterating the procedure, the impossibility to create bipartite entanglement in clusters by LOCCs holds at each step. When fewer than $k$ particles remain, we study bipartite entanglement in clusters of progressively decreasing size. When only $k-1$ particles are left, we compute  $E[\rho_{N}]\left(\mathcal{X}^{N-k+2}_{[1]}:\mathcal{X}^{N-k+2}_{[2]}\mathcal{X}^{N-k+2}_{[3]}...  \mathcal{X}^{N-k+2}_{[k-1]}\right)$, and so on. The last, $N-1$-th step is to quantify $E[\rho_N]\left(\mathcal{X}^{N-1}_{[1]}:\mathcal{X}^{N-1}_{[2]}\right)$.  \\Finally, we call $s_{N,k}(\rho_N)$ the sum of all bipartite entanglement terms for this specific sequence of particle labeling:
\begin{align}
\Big\{\, & \big(\mathcal{X}^1_{[1]}:\mathcal{X}^1_{[2]}...\mathcal{X}^1_{[k]}\big),\big(\mathcal{X}^2_{[1]}:\mathcal{X}^2_{[2]}...\mathcal{X}^2_{[k]}\big), \nonumber \\& ...,\big(\mathcal{X}^{N-k+2}_{[1]}:\mathcal{X}^{N-k+2}_{[2]}...\mathcal{X}^{N-k+2}_{[k-1]}\big),\nonumber \\&...,\big(\mathcal{X}^{N-1}_{[1]}:\mathcal{X}^{N-1}_{[2]}\big)
\Big\} 
\end{align}

Next, let us define the set  ${\cal S}_{N,k}$   of  sums $s_{N,k}$ associated to all possible sequences. 
If  and only if, for any $s_{N,k} \in {\cal S}_{N,k}$ of any $k$, {\it all}  bipartite entanglement terms are zero,  then the  global state $\rho_N$ belongs to the set of fully separable  states $\text{Sep}=\{\rho_{sep}:=\sum_i p_i \otimes_{\alpha=1}^N\rho_{[\alpha],i}\}$.  That is,  monitoring bipartite entanglement across bipartitions of size up to $k$ is sufficient to evaluate the operational cost of multipartite states. \\
Further, the result suggests that  the sum of the bipartite entanglement contributions is a quantifier of $k>2$-partite entanglement: \\
\begin{figure}
\includegraphics[width=0.47\textwidth,height=5cm]{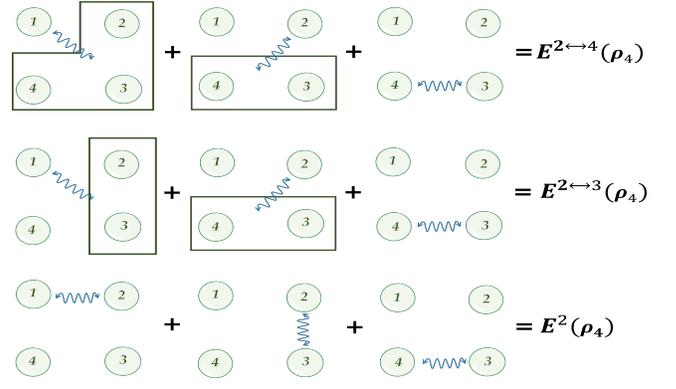}
    \caption{We  calculate  multipartite entanglement   in a 4-partite state $\rho_4$ with components labelled $1,2,3,4$, based on Eqs.~(\ref{eq1},\ref{eq2}). The nine pictures depict the partitions in which to calculate bipartite entanglement, for a given choice of particle labelling. When particles are inside a frame, they form a single cluster. For example, the upper left picture represents  $ E[\rho_4]\left(\mathcal{X}_{[1]}:\mathcal{X}_{[2]}\mathcal{X}_{[3]}\mathcal{X}_{[4]}\right)$.
The total entanglement  $E^{2\leftrightarrow 4}(\rho_4)$  is the sum of the three depicted terms in the top row,  maximized over all possible particle permutations. The (maximal) sum of the central  row  terms is $E^{2\leftrightarrow 3}(\rho_4)$. The difference between these two quantities  is $E^{4}(\rho_4)$.
Bipartite entanglement $E^2(\rho_4)$ is the maximal sum of the three terms $ E[\rho_4](\mathcal{X}_{[i]}:\mathcal{X}_{[j]})$ in the bottom row. Subtracting them to $E^{2\leftrightarrow 3}(\rho_4)$ gives the 3-partite entanglement $E^{3}(\rho_4)$.} 
\label{fig2}
\end{figure}
\noindent\textbf{Result 1 --} {\it The total amount of entanglement up to degree $k$,
i.e. the sum of entanglement of degrees $2,3,\ldots,k$ in $\rho_N$, is given by
\begin{align}
E^{2\leftrightarrow k}(\rho_N)
:=\max_{s_{N,k}\in {\cal S}_{N,k}}
s_{N,k}(\rho_N),
\label{eq1}
\end{align}
and the genuine $k$-partite entanglement  of $\rho_N$ is  
\begin{equation}\label{eq2}
    E^k(\rho_N):=E^{2\leftrightarrow k}(\rho_N)-E^{2\leftrightarrow k-1}(\rho_N).
\end{equation}}
We discuss the simplest case $N=3$, while in FIG.~\ref{fig2} we illustrate $N=4$.  One has 
\begin{align}
E^{2\leftrightarrow 3}(\rho_3)=&\max_{ijk=1,2,3}\{E[\rho_3]\left(\mathcal{X}_{[i]}:\mathcal{X}_{[j]}\mathcal{X}_{[k]}\right)\nonumber\\+&E[\rho_3]\left(\mathcal{X}_{[j]}:\mathcal{X}_{[k]}\right)\}\nonumber\\
 E^{2}(\rho_3)=&\max_{ijk=1,2,3}\{E[\rho_3]\left(\mathcal{X}_{[i]}:\mathcal{X}_{[j]}\right)\nonumber\\
 +&E[\rho_3]\left(\mathcal{X}_{[j]}:\mathcal{X}_{[k]}\right)\}.
\end{align}
The first term is the total entanglement, given by the sum of the 1 {\it vs} 2 particle entanglement and the bipartite entanglement between the remaining two particles, maximized over all particle permutations.  
 The second one is the maximal sum of bipartite entanglement terms, while $E^{3}(\rho_3)=E^{2\leftrightarrow 3}(\rho_3)-E^{2}(\rho_3)$  is the ``genuine'' $3$-partite entanglement. In general, when the system is invariant under particle exchange,
  $k$-partite entanglement  takes a compact form:
 \begin{align}
 E^{k}(\rho_N)=&(N-k+1)\{E[\rho_N](\mathcal{X}_{[1]}:\mathcal{X}_{[2]}\ldots\mathcal{X}_{[k]})\\\nonumber
 -& E[\rho_N](\mathcal{X}_{[1]}:\mathcal{X}_{[2]}\ldots\mathcal{X}_{[k-1]})\}.
\end{align} 

Next, we discuss the main properties of  $E^{2\leftrightarrow k}(\rho_N)$ and $E^{k}(\rho_N)$, which hold independently of the chosen  measure $E$. \\
The first important one is {\it faithfulness}. There is no entanglement of any degree if and only if the global state is in $\text{Sep}$: $E^{k}(\rho_N)=E^{2\leftrightarrow k}(\rho_N)=0, \forall k \iff \rho_N\in \text{Sep}$. More generally, if and only if the global state  is  a mixture of partially separable states
\begin{align}  
\sum_i p_i\rho_{{k_1,i}}\otimes\rho_{{k_2,i}}\otimes
\ldots\otimes\rho_{k_\alpha,i}, \, k_{1,2,\ldots,\alpha}\leq k,
\end{align}
 then $E^{l>k}(\rho_N)=0, E^{2\leftrightarrow N}(\rho_N)=E^{2\leftrightarrow k}(\rho_N)$. For example, a three-particle state $\rho_3=\sum_i p_i \rho_{2,i}\otimes \rho_{[3],i}$ does not have three-partite entanglement, if the particle $[3]$ is the same in any term of the decomposition. \\  
 
 Also, the measures defined in Eqs.~(\ref{eq1},\ref{eq2})  detect high-order entanglement  in the global state even when smaller partitions are in separable states. E.g., in the GHZ state of N qubits $\ket{\text{GHZ}_N}=(\ket{0}^{\otimes N}+\ket{1}^{\otimes N})/\sqrt{2}$ \cite{ghz}, one has  $E^{k}(\ket{\text{GHZ}_N})=0,\, \forall k<N,$ while $E^{N}(\ket{\text{GHZ}_N})\neq 0$. On the other hand,  only bipartite entanglement is found in  a product of Bell states: $E^k\left( \ket{\text{GHZ}_2}^{\otimes N/2}\right)=0,\,\forall k>2$, with $E^2\left( \ket{\text{GHZ}_2}^{\otimes N/2}\right)=N/2\, E^2( \ket{\text{GHZ}_2})$.
 \\
A second crucial constraint is invariance under single particle unitary transformations $U_{[i]}$. Since any bipartite  entanglement term in Eq.~(\ref{eq1}) is invariant, then both $E^{k}(\rho_N)$ and $E^{2\leftrightarrow k}(\rho_N)$ are unchanged.\\
Finally, we discuss monotonicity under  LOCCs. Any  $E[\rho_N](\mathcal{X}_{[i]}:\mathcal{X}_{[i+1]}\mathcal{X}_{[i+2]}\ldots\mathcal{X}_{[k]})$ is non-increasing under LOCCs performed on any pair of particles $\mathcal{X}_{[i]}\mathcal{X}_{[j]}$.  As $E^{2\leftrightarrow k}(\rho_N)$ is a sum  of these bipartite entanglement terms, it inherits monotonicity under two-particle LOCCs, for any $k$. Therefore, it is a full-fledged entanglement monotone.  Note that $E^{k}(\rho_N),$ for $k>2$, is a difference of monotone quantities (see Eq.~(\ref{eq2})), so it is not always decreasing under LOCCs. This is expected: given an initial state with multipartite entanglement, it is possible to increase bipartite entanglement by LOCCs \cite{conv}. Yet, the monotonicity of  $E^{2\leftrightarrow k}(\rho_N)$ dictates that, if entanglement of a specific degree $E^{l\leq k}(\rho_N)$ increases by LOCCs,  then $\sum_{m\neq l, m\leq k}E^m(\rho_N)$ decreases by, at  least,  the same amount of bits.\\

Further, we demonstrate  that this kind of multipartite entanglement determines the operational cost of  quantum state preparation:\\
\begin{table*}[t!h!]
    \centering
    \renewcommand{\arraystretch}{1.4}      
    \setlength{\tabcolsep}{7pt}            
    \begin{tabular}{|c|c|c|c|c|c|}
        \hline
        State 
        & $E_F^2$ & $E_F^3$ & $E_F^{3<k<N}$  &$E_F^{N}$ & $E_F^{2\leftrightarrow N}$ \\ 
        \hline
        \hline
       
        $\ket{\text{GHZ}_N}$       & 0    & 0    & 0  & 1  & 1\\
        \hline
           $\ket{\text{GHZ}_k}\otimes \ket{0}^{\otimes N-k}$    & 0    & 0    & 1             & 0 & 1\\
        \hline
       $\ket{\text{GHZ}_k} \otimes \ket{\text{GHZ}_{3}}\otimes \ket{0}^{\otimes N-k-3}$  & 0      & 1    &  1     &0   &2  \\
        \hline
         $\ket{\text{GHZ}_2}^{\otimes N/2}$ & N/2&0&0&0&N/2\\
         \hline
        $\ket{W_3}\otimes \ket{0}^{\otimes N-3}$    &   1.1  & 0.9   & 0&    0  &2\\
        \hline
        $\ket{W_3}\otimes \ket{\text{GHZ}_3}\otimes \ket{0}^{\otimes N-6}$    & 1.1    & 1.9   &0 &   0  &3\\
          \hline
        $\ket{W_N}$    & $(N-1)\,g(2,N)$  & \makecell{$(N-2)$\\$\{g(3,N)-$\\
        $g(2,N)\}$}   &\makecell{$(N-k+1)$\\$\{g(k,N)-$\\$g(k-1,N)\}$ }  & \makecell{ $g(N,N)-$\\$g(N-1,N)$}& \makecell{$\sum_{\alpha=2}^N g(\alpha,N) $}\\
        \hline
    \end{tabular}
    \caption{We compute the $k$-partite entanglement of formation $E_F^k$ and the  total entanglement of formation $E_F^{2\leftrightarrow N}$ (see Eqs.~(\ref{eq1},\ref{eq2})) in several  states. The entanglement ``vector'' $E_F^{k},  2\leq k\leq N,$ is different for each of those states, capturing the distinct features of their correlation structures. Because of Result 3,  it has a closed form for  $W_N$ states, which depends on the function $g(k,N)=h\left((1+\sqrt{1-4 (k-1)/N^2})/2\right)$, allowing to study the scaling of quantum correlations (see FIG.~\ref{plotsW}).}
    \label{tab1}
\end{table*}
{\bf Result 2 --} {\it If $E$ is a convex roof entanglement monotone,  then $E^k(\rho_N)\neq 0$  only if preparing $\rho_N$ from $\text{Sep}$ requires quantum communication between at least  $k-1$ particle pairs. \\ 
If $E$ is a generic bipartite entanglement measure, the claim holds for pure state transformations from $\otimes_{\alpha=1}^N\psi_{[\alpha]}$ to  $\ket{\psi}_N$.}
\begin{proof} Convex roof  measures of bipartite entanglement in a state $\rho_2$ are  defined as
\begin{eqnarray}\label{convex}
E_{cr}[\rho_{2}]({\mathcal X}_{[1]}:{\mathcal X}_{[2]}):=\min_{\{p_i,\,\rho_{2,i}\}}\sum_i p_i f(\rho_{2,i}),
\end{eqnarray}
where the minimization is over all the state decompositions $\rho_2=\sum_ip_i\rho_{2,i}$, and $f$ is a function that is invariant under local unitaries and non-increasing on average under single-particle measurements \cite{croof1,croof2}. The entanglement of formation $E_F$ is a notable example \cite{locc1}. It is defined by constraining $\rho_{2,i}$  to be pure states and taking $f$ to be the entropy of single-particle marginal states, which is the entanglement of each term in the  decomposition. 
 A two-particle entangling operation $\Phi_{[1][2]}$, transforms a
 $\rho_{sep}$ into 
\begin{align}
\tilde \rho_N=&\Phi_{[1][2]}(\rho_{sep}) 
=\sum_i p_i \rho_{2,i}\otimes\left( \otimes_{\alpha=3}^N \rho_{[\alpha],i}\right).
\end{align}
The convexity of $E_{cr}$ implies that
\begin{align}
&E_{cr}[\tilde \rho_N]\left({\mathcal X}_{[1]}:{\mathcal X}_{[2]}\right)\leq  E_{cr}[\tilde \rho_N]\left({\mathcal X}_{[1]}:{\mathcal X}_{[2]}{\mathcal X}_{[3]}\ldots {\mathcal X}_{[k]}\right)\nonumber\\
\leq& \sum_i p_i E_{cr}\left[\rho_{2,i}\otimes\left( \otimes_{\alpha=3}^N \rho_{[\alpha],i}\right)\right]\left({\mathcal X}_{[1]}:{\mathcal X}_{[2]}{\mathcal X}_{[3]}\ldots {\mathcal X}_{[k]}\right)\nonumber\\
=& \sum_i p_i E_{cr}[\rho_{2,i}]\left({\mathcal X}_{[1]}:{\mathcal X}_{[2]}\right)
\leq \sum_{i,j^i} p_i q_{j^i} f(\sigma_{2,j^i})\nonumber\\
=& \sum_{l} r_l f(\sigma_{2,l}),
\end{align}
where $\rho_{2,i}=\sum_{j^i}q_{j^i} \sigma_{2,j^i}$. The result applies to any decomposition $\{r_l, \sigma_{2,l}\}$ of $\tilde{\rho}_2$, including the one  that solves the minimization in Eq.~(\ref{convex}). Hence, one has
\begin{align}
E_{cr}[\tilde \rho_N]\left({\mathcal X}_{[1]}:{\mathcal X}_{[2]}\right)=&  E_{cr}[\tilde \rho_N]\left({\mathcal X}_{[1]}:{\mathcal X}_{[2]}{\mathcal X}_{[3]}\ldots {\mathcal X}_{[k]}\right)\nonumber\\
&\Rightarrow E_{cr}^k(\tilde \rho_N)=  0,\,\forall\, k>2,
\end{align}
which means that a single two-particle operation can create only bipartite entanglement from a state in $\text{Sep}$. Iterating the proof  demonstrates that one needs at least $k-1$ entangling operations to prepare $k$-particle entanglement. Note that, as any $\rho_{sep}$ can be written as a mixture of pure states, the result applies to the entanglement of formation. Also, the restriction to two-particle operations does not imply loss of generality, as any quantum operation can be synthesized with arbitrary precision by  two-particle and single-particle gates \cite{NielsenChuang,kitaev,divinc}.\\
If the initial state is a product of single particle pure states $\ket{\psi_{N}}=\otimes_{\alpha=1}^N\ket{\psi_{[\alpha]}}$,  any two-particle operation   increases the maximal degree  of entanglement by at most one. Thus, for any choice of $E$, one needs  at least $k-1$ entangling operations to create $k$-partite entanglement. \end{proof}
This definition of multipartite entanglement is  different from the extensively studied notions based on  separability and producibility.  Specifically,
 an $N$\textendash partite pure state $\ket{\psi_N}$ is called $k$\textendash separable if it is a product of $k$ terms:
\begin{equation}
   \ket{\psi_1}_{[1]\ldots[\alpha_1]}\otimes\ket{\psi_2}_{[\alpha_1+1]\ldots[\alpha_2]}\otimes \dots \otimes\ket{\psi_k}_{[\alpha_{k-1}+1]\ldots[\alpha_k]}.
\end{equation}
Extended to mixed states, the definition classifies quantum states by the number of separable partitions \cite{sep1,sep2}.\\ 
\begin{figure}[h!]
\includegraphics[width=.47\textwidth,height=5.5cm]{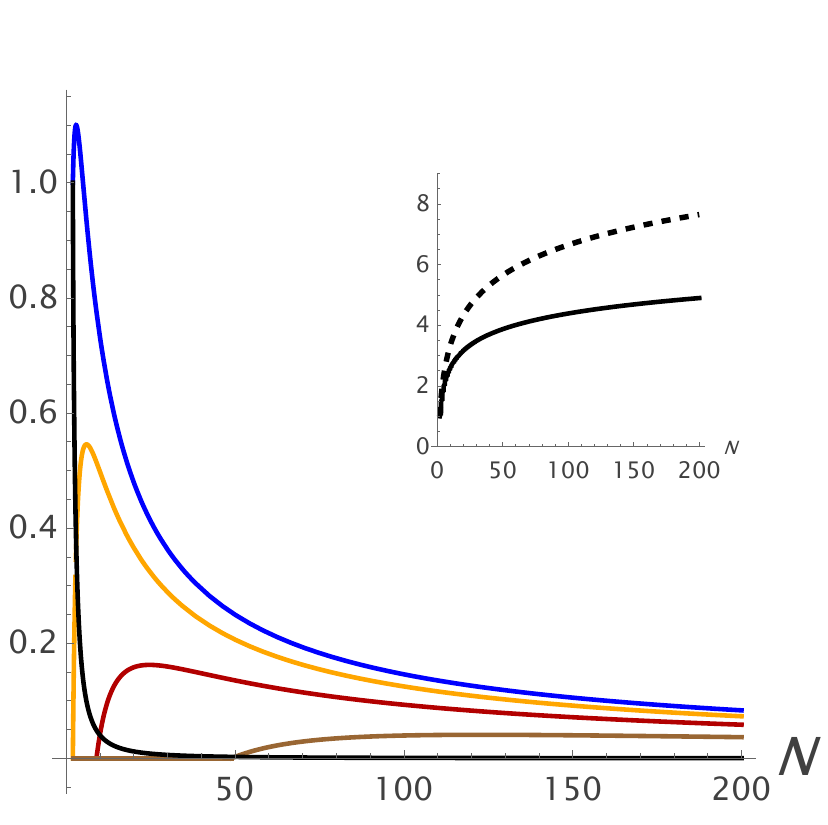}
\caption{We study how  multipartite entanglement scales with system size in $W_N$ states. We plot $E^2_F$ (blue line), $E^3_F$ (orange),  $E^{10}_F$ (red), $E^{50}_F$ (brown) $E^N_F$ (black). By increasing $N$, $k$-partite entanglement exponentially decreases (for any $k$), as  the  similarity between $W_N$ and its partitions increases with the number of particles. In the inset, we plot the total entanglement $E^{2\leftrightarrow N}_F$ (black continuous line), which is upper bounded by $\log_2 N$ (dashed line). The analytical expressions of these quantities are given in TABLE \ref{tab1}.}
\label{plotsW}
\end{figure}
While an important feature, separability is unrelated to the experimental cost of multipartite entanglement.  For example, the state $\ket{\text{GHZ}_{N-1}}\otimes \ket{0}$ remains biseparable for any $N$, while its preparation demands quantum circuits linearly increasing with the particle number.\\
Producibility instead counts the size of clusters in which each particle is entangled with the rest of the cluster. An $N$\textendash partite pure state  is \textit{$k$\textendash producible} if expressed by product of at most $k$-particle states:
\begin{align}\label{pure_producib}
    &\ket{\psi_1}_{[1]\ldots[\alpha_1]}\otimes \ket{\psi_2}_{[\alpha_1+1]\ldots[\alpha_2]}\otimes \dots \otimes \ket{\psi_m}_{[\alpha_{m-1}+1]\ldots[\alpha_m]}, \nonumber\\&\alpha_{1,2,\ldots,m}\leq k.
\end{align}
The concept can  be extended to mixed states \cite{prod,szalay0,szalay}.\\ 
Producibility does not capture the experimental cost of entanglement. For example, production of  a product of $N/2$ Bell states is achieved by $N/2$ gates, while being $2$-producible, for any $N$. Also, engineering the three-qubit, bi-producible mixed states $\sum_i p_i \rho_{[1][2],i}\otimes\rho_{[3]}+\rho_{[1],i}\otimes\rho_{[2][3]}$ needs entangling gates between two particle pairs, just like the three-producible state $\ket{\text{GHZ}_3}$. Indeed, one has $E^3\neq 0$ in both states. For the same reason, the size of the largest cluster in Eq.~(\ref{pure_producib}), called entanglement depth \cite{depth}, does not quantify the cost of multipartite entanglement.  The same argument applies to the Schmidt number, as it is not additive under tensor products of a state copies \cite{schmidt}. \\

We explicitly calculate the newly defined  $k$-partite entanglement of formation $E_F^k(\rho_N)$ and  the related total entanglement of formation $E_F^{2\leftrightarrow N}(\rho_N)=\sum_{k=2}^N E_F^k(\rho_{N})$  in relevant case studies 
(TABLE~\ref{tab1}). \\
As expected, for the maximally entangled $\text{GHZ}_N$ states, the quantities signal that there is only the highest degree of entanglement $E^N_F$, also called ``genuine multipartite entanglement'' \cite{rev,gme1,gme2,gme3,gme4,multient_termo,two_par_fam_multient}. Remarkably, we found a closed expression of this quantity for  the W states: 
\begin{align}\ket{\text{W}_N}= \frac{1}{\sqrt{N}}\sum_{i=1}^N\ket{0}_{[1]}\otimes\ldots\otimes \ket{1}_{[i]}\otimes \ldots\otimes\ket{0}_{[N]}.
\end{align}
 In particular:\\
{\bf Result 3 --} {\it    Consider a $\mathcal{X}_k$ cluster of  $N$ qubits in the state $W_N$. 
    The bipartite entanglement of formation in its state $\rho_{\text{W},k}$ between  $m$ qubits and the remaining $k-m$ qubits in $\mathcal{X}_k$ is
\begin{align}
        E_F[\rho_{W,k}](\mathcal{X}_{[1]}\ldots \mathcal{X}_{[m]}:\mathcal{X}_{[m+1]}\ldots \mathcal{X}_{[k]})
        =&\nonumber\\
        h\!\left(\frac{1+\sqrt{1-\frac{4m(k-m)}{N^2}}}{2}\right),
\end{align}
  where $h(x)=-x\log x -(1-x)\log(1-x)$. \\
Therefore, $E_F^k[\ket{W_N}]$ is analytically computable.}\\
The proof is reported in \cite{supmat}. The result is surprising,  as the bipartite entanglement of formation has a generic closed expression only for two-qubit mixed states \cite{wootters}, and an analytical upper bound for two-qudit states \cite{osborne}. \\
The $W_N$ states are interesting because they display multipartite entanglement of any degree $k$. That is, they are the simplest class of multipartite states with a non-trivial entanglement structure, see FIG.~\ref{plotsW}.  Also, they have important applications in quantum information processing \cite{entapp1,w1,w2,w3}. Their quantum correlations are indeed more robust than those of the GHZ states, as their clusters $W_k$ are still entangled. \\

{\it Conclusion --}
We introduced a method to quantify the operational cost of creating  multipartite entanglement. 
The result opens several areas of investigation.  The construction of multipartite entanglement measures from bipartite entanglement can be generalized to  other kinds of quantum correlations, e.g.,  non-locality, steering, and discord \cite{steering,discord,nonlocality}. Also, it may enable the experimental estimation of  multipartite entanglement, rather than its mere detection \cite{exp1,exp2,exp3,exp4,exp5,exp6,exp7,exp8,exp9}, which is essential to certify  the success of quantum state preparation protocols. Indeed, based on Eqs.~(\ref{eq1},\ref{eq2}), directly observable quantifiers of bipartite entanglement can be readily applied to detect multipartite entanglement. 
Moreover, as bipartite entanglement has become a crucial tool to investigate condensed matter and high-energy systems, we expect multipartite entanglement to be a signature of critical properties in many-body quantum dynamics \cite{enthigh,entcond}.
The interplay between multipartite entanglement and circuit complexity \cite{eisert,me,Girolami}, and more generally the quantification of resource cost of quantum information processing, also deserves further investigation.


\section*{Acknowledgments}
This research was supported by the Italian Ministry of Research, grant number MUR-PRIN2022, Contract Number NEThEQS (2022B9P8LN).

\clearpage
\onecolumngrid
\renewcommand{\bibnumfmt}[1]{[A#1]}
 
\renewcommand{\citenumfont}[1]{{A#1}}

\setcounter{page}{1}
\setcounter{equation}{0}

\appendix*

\section{{\large Quantifying the Operational Cost of Multipartite Entanglement
}}

\subsection*{SUPPLEMENTARY MATERIAL} 
 


 \subsection*{Demonstration of Result 3 of main text} 
 We calculate the closed form of the $k$-partite entanglement of formation of $W_N$ states. Since the entire procedure is based on the computation of bipartite entanglement, it is sufficient to analytically derive the bipartite entanglement of formation between any two clusters. Specifically, given an $N$-particle system in the $W_N$ state, if $\rho_{W,k}$ is the state obtained after tracing out $N-k$ qubits, the entanglement between $m$ of these qubits and the remaining $k-m$ is given by
    \[
        E_F[\rho_{W,k}]\left(\mathcal{X}_{[1]}\ldots \mathcal{X}_{[m]}:\mathcal{X}_{[m+1]}\ldots \mathcal{X}_{[k]}\right)
        =h\!\left(\frac{1+\sqrt{1-\frac{4m(k-m)}{N^2}}}{2}\right),
    \]
    where $h$ denotes the classical binary entropy.

\begin{proof*}

Let
\begin{equation*}
\ket{W_N}=\frac{1}{\sqrt{N}}\sum_{i=1}^N \ket{0\cdots 1_{[i]} \cdots 0}
\end{equation*}
be the $N$-qubit $W$ state. Tracing out $N-k$ qubits yields
\begin{equation*}
\rho_{W,k}=\frac{k}{N}\ket{W_k}\!\bra{W_k}
+\left(1-\frac{k}{N}\right)\ket{0^k}\!\bra{0^k}.
\end{equation*}

We partition the $k$ qubits into two clusters $A$ and $B$ containing $m$ and $k-m$ qubits, respectively. Across this bipartition, the state $\ket{W_k}$ admits the following decomposition:
\begin{align*}
\ket{W_k}
&=\frac{1}{\sqrt{k}}\sum_{i=1}^k \ket{0\cdots 1_{[i]} \cdots 0}\\
&=\frac{1}{\sqrt{k}}\Bigl(
\sum_{i=1}^{m}\ket{0\cdots 1_{[i]} \cdots 0}_A\ket{0^{k-m}}_B \\
&\quad+\sum_{i=m+1}^{k}\ket{0^{m}}_A\ket{0\cdots 1_{[i]} \cdots 0}_B
\Bigr)\\
&=\frac{1}{\sqrt{k}}\Bigl(\sum_{i=1}^{m}\ket{0\cdots 1_{[i]} \cdots 0}_A\Bigr)\ket{0^{k-m}}_B\\
&\quad+\frac{1}{\sqrt{k}}\ket{0^{m}}_A\Bigl(\sum_{i=1}^{k-m}\ket{0\cdots 1_{[i]} \cdots 0}_B\Bigr).
\end{align*}
Using $\sum_{i=1}^{m}\ket{0\cdots 1_{[i]} \cdots 0}_A=\sqrt{m}\,\ket{W_m}_A$ and
$\sum_{i=1}^{k-m}\ket{0\cdots 1_{[i]} \cdots 0}_B=\sqrt{k-m}\,\ket{W_{k-m}}_B$, we obtain
\begin{equation*}
\ket{W_k}
=\sqrt{\frac{m}{k}}\ket{W_m}_A\ket{0^{k-m}}_B
+\sqrt{\frac{k-m}{k}}\ket{0^m}_A\ket{W_{k-m}}_B
\end{equation*}

The states appearing in this decomposition are orthonormal. The product state $\ket{0^k}=\ket{0^m}_A\ket{0^{k-m}}_B$ involves exactly the same local vectors $\ket{0^m}$ and $\ket{0^{k-m}}$ already present in the decomposition of $\ket{W_k}$. Therefore, both terms in $\rho_{W,k}$ have support on the same local subspaces. It follows that the reduced density operators of $\rho_{W,k}$ satisfy
\begin{equation*}
    \operatorname{supp}(\rho_A)=\mathrm{span}\{\ket{0^m},\ket{W_m}\} := \mathcal S_A
\end{equation*}
\begin{equation*}
    \operatorname{supp}(\rho_B)=\mathrm{span}\{\ket{0^{k-m}},\ket{W_{k-m}}\} := \mathcal S_B
\end{equation*}

We now define two local isometries $V_A:\mathbb{C}^2\to(\mathbb{C}^2)^{\otimes m}$ and
$V_B:\mathbb{C}^2\to(\mathbb{C}^2)^{\otimes(k-m)}$ such that
\begin{equation*}
    V_A\ket{0}=\ket{0^m},\quad V_A\ket{1}=\ket{W_m}
\end{equation*}
\begin{equation*}
    V_B\ket{0}=\ket{0^{k-m}},\quad V_B\ket{1}=\ket{W_{k-m}}
\end{equation*}
Because $\rho_{W,k}$ has support only on $\mathcal S_A \otimes \mathcal S_B$, it can be written as
\begin{equation}
\rho_{W,k}
=
\sum_{i,j,k,l=0}^{1}
\rho_{ij,kl}\,
\ket{a_i b_j}\!\bra{a_k b_l}
\end{equation}
with complex coefficients $\rho_{ij,kl}$. Using the definition of the local isometries, we have
\begin{equation*}
\ket{a_i b_j}=(V_A\otimes V_B)\ket{ij},
\qquad
\bra{a_k b_l}=\bra{kl}(V_A^\dagger\otimes V_B^\dagger)
\end{equation*}
Substituting these identities term by term into the expansion of $\rho_{W,k}$ gives
\begin{align}
\rho_{W,k}
&=
\sum_{i,j,k,l=0}^{1}
\rho_{ij,kl}\,
(V_A\otimes V_B)\ket{ij}\bra{kl}(V_A^\dagger\otimes V_B^\dagger)
\nonumber\\
&=
(V_A\otimes V_B)
\left(
\sum_{i,j,k,l=0}^{1}
\rho_{ij,kl}\,
\ket{ij}\bra{kl}
\right)
(V_A^\dagger\otimes V_B^\dagger)
\end{align}

We now note that the isometries $V_A$ and $V_B$ are not invertible on the full Hilbert spaces, but their adjoints act as inverses on their images. Since
$V_A^\dagger V_A=\mathbb I_2$ and $V_B^\dagger V_B=\mathbb I_2$, we have
\begin{equation*}
(V_A^\dagger\otimes V_B^\dagger)(V_A\otimes V_B)=\mathbb I_{2\otimes2}
\end{equation*}
Because $\rho_{W,k}$ has support entirely contained in
$\mathrm{Im}(V_A\otimes V_B)=\mathcal S_A\otimes\mathcal S_B$, the operator
$V_A^\dagger\otimes V_B^\dagger$ acts as a true inverse on $\rho_{W,k}$. Therefore, the two-qubit density matrix
\begin{equation}
\rho_{2\otimes2}=(V_A^\dagger\otimes V_B^\dagger)\,\rho_{W,k}\,(V_A\otimes V_B)
\end{equation}
is uniquely defined and satisfies
\begin{equation}
\rho_{W,k}=(V_A\otimes V_B)\,\rho_{2\otimes2}\,(V_A^\dagger\otimes V_B^\dagger).
\end{equation}

The existence of this reduction relies precisely on the two-dimensionality of the local supports. Explicitly,
\begin{equation*}
(V_A^\dagger\otimes V_B^\dagger)\ket{0^k}=\ket{00},
\end{equation*}
\begin{equation*}
(V_A^\dagger\otimes V_B^\dagger)\ket{W_k}
=\sqrt{\frac{m}{k}}\ket{10}+\sqrt{\frac{k-m}{k}}\ket{01}
\end{equation*}
so that
\begin{equation}
\rho_{2\otimes2}
=\frac{k}{N}\ket{\psi_{m,k}}\!\bra{\psi_{m,k}}
+\left(1-\frac{k}{N}\right)\ket{00}\!\bra{00}
\end{equation}
where
\begin{equation}
\ket{\psi_{m,k}}=\sqrt{\frac{m}{k}}\ket{10}+\sqrt{\frac{k-m}{k}}\ket{01}
\end{equation}

Since these local isometries are LOCCs, the entanglement between the two clusters in $\rho_{2\otimes2}$ is the same as the one between the selected $m$ qubits and the remaining $k-m$ qubits in $\rho_{W,k}$. Thus, from now on we restrict our attention to $\rho_{2\otimes2}$. The concurrence of this two-qubit state can be computed using the result in \cite{Wootters_conc,eberly} and is given by $\frac{2}{N}\sqrt{m(k-m)}$. 
Finally, the corresponding entanglement of formation  is
\begin{equation}\label{a7}
E_F[\rho_{W,k}](\mathcal{X}_{[1]}\ldots \mathcal{X}_{[m]}:\mathcal{X}_{[m+1]}\ldots \mathcal{X}_{[k]})
=h\!\left(\frac{1+\sqrt{1-\frac{4m(k-m)}{N^2}}}{2}\right).
\end{equation}
\end{proof*}

This result  allows us to compute the entanglement across any possible partition of the $W$ state. Note that  it is consistent with the upper bound found for two-qudit states in \cite{osbornesupp}. In this case, the bound is exactly saturated, matching  the entanglement of formation between two clusters of the $W_N$ state.\\
 
In conclusion, the $k$-partite entanglement of formation is a function of the bipartite entanglement between one qubit  and a varying number of qubits. Hence, by fixing $m=1$, the Result 3 of the main text, i.e., Eq.~(\ref{a7}),  enables one to analytically reconstruct the whole entanglement vector $E_F^k(\ket{W_N}), 2\leq k\leq N$.


\begin{thebibliography}{99}

 
\bibitem{Jozsa-Linden} R. Jozsa and N. Linden, On the role of entanglement in quantum-computational speed-up, Proc. Royal Soc.
 {\bf 459}, pp. 2011–2032 (2003).
 
 \bibitem{vidal}G. Vidal, Efficient classical simulation of slightly entangled quantum computations, Phys. Rev. Lett. {\bf 91}, 147902 (2003).
 
\bibitem{rev}R. Horodecki, P. Horodecki, M. Horodecki, and K. Horodecki,
Quantum entanglement, Rev. Mod. Phys. {\bf 81}, 865-942 (2009).

\bibitem{entapp1}G. Toth, Multipartite entanglement and high-precision metrology,
Phys. Rev. A {\bf 85}, 022322 (2012).
\bibitem{comm}Y. Yeo and W. K. Chua, Teleportation and Dense Coding with Genuine Multipartite Entanglement, 
Phys. Rev. Lett. {\bf 96}, 060502 (2006).

 \bibitem{book} I. Bengtsson and K. \.{Z}yczkowski, \textit{Geometry of Quantum States: An Introduction to Quantum Entanglement} (Cambridge University Press, 2007).
 
 

\bibitem{locc1}
C. H. Bennett,  D. P. DiVincenzo, J. A. Smolin, and W. K. Wootters, Mixed-state entanglement and quantum error correction, Phys. Rev. A {\bf 54}, 3824 (1996).

\bibitem{locc3}
E. Chitambar, D. Leung, L. Mancinska, M. Ozols, and A. Winter, Everything You Always Wanted to Know About LOCC (But Were Afraid to Ask), 	Commun. Math. Phys. {\bf 328}, 1, 303-326 (2014). 


\bibitem{sep1} W. Dür, J. I. Cirac, R. Tarrach, Separability and distillability of multiparticle quantum systems, Phys. Rev. Lett. {\bf 83}, 3562-3565 (1999).
 
 \bibitem{sep2}W. Dür, G. Vidal, and J. I. Cirac, Three qubits can be entangled in two
inequivalent ways, Phys. Rev. A {\bf 62}, 062314 (2000).
\bibitem{sep3}
F. Verstraete, J. Dehaene, B. De Moor, and H. Verschelde, Four qubits can be entangled in nine different ways, Phys. Rev. A {\bf 65}, 052112 (2002).

\bibitem{sep4}O. Guehne and M. Seevinck, Separability criteria for genuine multiparticle entanglement, New J. Phys. {\bf 12}, 053002 (2010).

\bibitem{sep5}A. Acín, D. Bruss, M. Lewenstein, A. Sanpera,  Classification of Mixed Three-Qubit States, Phys. Rev. Lett. {\bf 87}, 040401 (2001).



\bibitem{sep6}
J. Eisert and H. J. Briegel, Schmidt measure as a tool for quantifying multiparticle entanglement, Phys. Rev. A {\bf 64}, 022306
(2001). 



\bibitem{szalay0}
S. Szalay,
k-stretchability of entanglement, and the duality of k-separability and k-producibility, 
Quantum 3, {\bf 204} (2019).

\bibitem{prod}O. Gühne, G. Tóth, and H. J. Briegel, Multipartite entanglement in spin chains, New J. Phys. {\bf 7}, 229 (2005).

\bibitem{szalay}S. Szalay. Multipartite entanglement measures. Phys. Rev. A {\bf 92}, 042329 (2015).
\bibitem{prod2}S. Wölk and O. Gühne, Characterizing the width of entanglement, New Journal of Physics {\bf 18}, 123024 (2016).







\bibitem{entmeasures1}V. Vedral, M. B. Plenio, M. A. Rippin, and P. L. Knight, Quantifying Entanglement,
Phys. Rev. Lett. {\bf 78}, 2275 (1997).






\bibitem{entmeasures2}A. Peres, Separability Criterion for Density Matrices, Phys. Rev. Lett. {\bf 77}, 1413 (1996).
\bibitem{entmeasures3}
V. Coffman, J. Kundu, and W. K. Wootters,
Distributed Entanglement, Phys. Rev. A {\bf 61}, 052306 (2000).
\bibitem{entmeasures4}
S. Xie and J. H. Eberly, Triangle Measure of Tripartite Entanglement,
Phys. Rev. Lett. {\bf 127}, 040403 (2021).
\bibitem{entmeasures5}
C. Eltschka and J. Siewert,
Entanglement of Three-Qubit Greenberger-Horne-Zeilinger–Symmetric States,
Phys. Rev. Lett. {\bf 108}, 020502 (2012).
\bibitem{entmeasures6}
Measure of multipartite entanglement with computable lower bounds
Y. Hong, T. Gao, and F. Yan,
Phys. Rev. A {\bf 86}, 062323 (2012).
\bibitem{entmeasures7}
 Y.-J. Luo, X. Leng, and C. Zhang,
 Genuine multipartite entanglement verification with convolutional neural networks,
Phys. Rev. A {\bf 110}, 042412 (2024).
\bibitem{entmeasures8}
R. K. Malla, A. Weichselbaum, T.-C. Wei, and R. M. Konik,
Detecting Multipartite Entanglement Patterns Using Single-Particle Green’s Functions,
Phys. Rev. Lett. {\bf 133}, 260202 (2024).
\bibitem{entmeasures9}F. Shi, L. Chen, G. Chiribella, and Q. Zhao, Entanglement Detection Length of Multipartite Quantum States,
Phys. Rev. Lett. {\bf134}, 050201 (2025). 
\bibitem{entmeasures10}S. Mukherjee, B. Mallick, S. Gopalkrishna Naik, A. G. Maity, and A. S. Majumdar, Detecting genuine multipartite entanglement using moments of positive maps,
Phys. Rev. A {\bf112}, 062428 (2025).
\bibitem{entmeasures11}
I. Biswas, A. Bhunia, S. Bera, I. Chattopadhyay, and Debasis Sarkar, Entanglement of assistance as a measure of multiparty entanglement,
Phys. Rev. A {\bf 111}, 032423 (2025).
\bibitem{entmeasures12}
H. Li, T. Gao, F. Yan, Parameterized multipartite entanglement measures, Phys. Rev. A {\bf 109}, 012213 (2024).

  \bibitem{entmeasures13}
C. Han, Y. Meir, and E. Sela,
Realistic protocol to measure entanglement at finite temperatures,
	Phys. Rev. Lett. 130, 136201 (2023).

\bibitem{ghz}D. M. Greenberger, M. A. Horne, and A. Zeilinger, Multiparticle
Interferometry and the Superposition Principle, Phys. Today {\bf 46}, 22–29 (1993).



\bibitem{conv}E. Chitambar, R. Duan, and Y. Shi, Multipartite-to-bipartite entanglement transformations and polynomial identity testing,
Phys. Rev. A {\bf 81}, 052310 (2010).
\bibitem{croof1}
G. A. Paz-Silva and J. H. Reina, Total correlations as fully additive entanglement monotones, J. Phys. A: Math. Theor. {\bf 42}, 055306 (2009).
\bibitem{croof2}D. Yang, K. Horodecki, M. Horodecki, P. Horodecki, J. Oppenheim, and Wei Song,
Squashed entanglement for multipartite states and entanglement measures based on the mixed convex roof,
 IEEE Trans. Inf. Theory {\bf 55}, 3375 (2009).


\bibitem{NielsenChuang}M. A. Nielsen and I. L. Chuang, \textit{Quantum Computation and Quantum Information: 10th Anniversary Edition} (Cambridge University Press, 2011).
 
 \bibitem{kitaev}A. Y. Kitaev. Quantum computations: algorithms and error correction. Russ. Math. Surv.,
{\bf 52}, 1191–1249, (1997).

 \bibitem{divinc}D. P. Divincenzo, Two-Bit Gates are Universal for Quantum Computation,	Phys. Rev. A {\bf 51}, 1015 (1995).
 
 
 

\bibitem{depth}Anders S. Sørensen and K. Mølmer, Entanglement and extreme spin squeezing, Phys. Rev. Lett. {\bf 86}, 4431 (2001).


\bibitem{schmidt}Q. Yue and E. Chitambar, The zero-error entanglement cost is highly non-additive, 
J. Math. Phys. {\bf 60}, 112204 (2019).


\bibitem{gme3}
J. L. Beckey, N. Gigena, P. J. Coles, and M. Cerezo, Computable and Operationally Meaningful Multipartite Entanglement Measures, Phys. Rev. Lett. {\bf 127}, 140501 (2021).
 

\bibitem{gme2}J. D. Bancal, N. Brunner, N. Gisin, and Y. C. Liang, Detecting Genuine Multipartite Quantum Nonlocality: A Simple Approach and Generalization to Arbitrary Dimensions, Phys. Rev. Lett. {\bf 106}, 020405 (2011).

\bibitem{gme1}M. Huber, F. Mintert, A. Gabriel, B. C. Hiesmayr, Detection of high-dimensional genuine multi-partite entanglement of mixed states, Phys. Rev. Lett. {\bf 104}, 210501 (2010).
\bibitem{gme4}
M. Huber and J. I. de Vicente, Structure of Multidimensional Entanglement in Multipartite Systems,
Phys. Rev. Lett. {\bf 110}, 030501 (2013). 
\bibitem{multient_termo}Bai, Chen-Ming, and Yu Luo. "Multipartite entanglement measures based on the thermodynamic framework." Physical Review A 112.3 (2025): 032424.

\bibitem{two_par_fam_multient}Luo, Yu, et al. "Family of two-parameter multipartite entanglement measures." arXiv preprint arXiv:2511.09415 (2025).
 



 \bibitem{supmat}See Supplementary  Material.

\bibitem{wootters}W. K. Wootters, Entanglement of formation of an arbitrary state of two qubits, Physical Review Letters {\bf 80}, 2245 (1998).
\bibitem{osborne}T. J. Osborne, 
Entanglement for rank-2 mixed states, 
Phys. Rev. A {\bf 72}, 022309 (2005). 






 

\bibitem{w1}P. Agrawal and A. Pati, Perfect teleportation and superdense coding
with W states, Phys. Rev. A {\bf 74}, 062320 (2006).

\bibitem{w2}M. D. Lukin and M. Fleischhauer, Quantum memory for photons: Dark-state polaritons. Phys. Rev. A. {\bf 65}, 022314 (2002).


\bibitem{w3}T. Tashima, T. Wakatsuki, Ş. K. Özdemir, T. Yamamoto, M. Koashi, and N. Imoto, Local transformation of two einstein-podolsky-rosen photon pairs into a three-photon W state, Phys. Rev. Lett. {\bf 102}, 130502 (2009).


\bibitem{steering}
R. Uola, A. C. S. Costa, H. C. Nguyen, and O. Gühne, Quantum Steering, 	Rev. Mod. Phys. {\bf 92}, 15001 (2020).

\bibitem{discord}
K. Modi, A. Brodutch, H. Cable, T. Paterek, and V. Vedral, The classical-quantum boundary for correlations: Discord and related measures, 
Rev. Mod. Phys. {\bf 84}, 1655 (2012).
\bibitem{nonlocality}N. Brunner, D. Cavalcanti, S. Pironio, V. Scarani, and S. Wehner, Bell nonlocality, Rev. Mod. Phys. {\bf 86}, 419 (2014).  


 
\bibitem{exp1}O. Gühne and G. Tóth, Entanglement detection, Physics Reports
{\bf 474}, 1–75 (2009).
\bibitem{exp2}P. Hyllus, W. Laskowski, R.
Krischek, C. Schwemmer, W. Wieczorek, H. Weinfurter, L. Pezzé, and A. Smerzi, Fisher information and multiparticle entanglement. Phys. Rev. A {\bf 85}, 022321,
2012.
\bibitem{exp3}Z. Qin, M. Gessner, Z.
Ren, X. Deng, D. Han, W. Li, X. Su, A. Smerzi, and
K. Peng, Characterizing the multipartite
continuous-variable entanglement structure from
squeezing coefficients and the fisher information.
NPJ Quant Inf. {\bf 5}, 3 (2019).
 
\bibitem{exp4}
M. Huber, F. Mintert, A. Gabriel, and B. C. Hiesmayr, Detection of High-Dimensional Genuine Multipartite Entanglement of Mixed States,
Phys. Rev. Lett. {\bf 104}, 210501 (2010).  

\bibitem{exp5}B. Lücke, J. Peise, G. Vitagliano, J. Arlt, L. Santos, G. Tóth, and C. Klempt. Detecting multiparticle entanglement of Dicke states. Phys. Rev. Lett. {\bf 112}, 155304 (2014).
\bibitem{exp6}D. Girolami and B. Yadin, Witnessing Multipartite Entanglement by Detecting Asymmetry, Entropy {\bf 19}, 124 (2017). https://doi.org/10.3390/e19030124

\bibitem{exp7}V. Saggio, A. Dimić, C. Greganti, L. A. Rozema, P. Walther, and B. Dakić, Experimental few-copy multipartite entanglement detection,
Nature Phys. {\bf 15}, 935–940 (2019).

\bibitem{exp8}H. Cao, S. Morelli, L. A. Rozema, C. Zhang, A. Tavakoli, and P. Walther, Genuine Multipartite Entanglement Detection with Imperfect Measurements: Concept and Experiment, Phys. Rev. Lett. {\bf 133}, 150201 (2024).

 \bibitem{exp9} Y.-Q. Zou, L.-N. Wu, Q. Liu, X.-Y. Luo, S.-F. Guo, J.-H. Cao, M. K. Tey, and L. You, Beating the classical precision limit with spin-1 Dicke states of more than 10,000 atoms, Proc.  Nat. Ac.  Sc. USA {\bf 115}, 6381–6385 (2018).
\bibitem{enthigh}
O. K. Baker, Quantum Entanglement in High Energy Physics, https://doi.org/10.5772/intechopen.111219 (IntechOpen, 2024).
 
\bibitem{entcond}N. Laflorencie, Quantum entanglement in condensed matter systems,  Phys. Rep. {\bf 643}, 1-59 (2016).

\bibitem{me}
D. Girolami and F. Anzà, Quantifying the difference between many-body quantum states, 
Phys. Rev. Lett. {\bf 126}, 170502 (2021).
\bibitem{Girolami}D. Girolami, T. Tufarelli, and C. E. Susa, Quantifying
Genuine Multipartite Correlations and their Pattern Complexity, Physical
Review Letters {\bf 119},  140505(2017).

 

\bibitem{eisert}J. Eisert, Entangling power and quantum circuit complexity, Phys. Rev. Lett. {\bf 127}, 020501 (2021)

 





 





















\end{thebibliography}

\begin{thebibliography}{99}



\bibitem{Wootters_conc}W. K. Wootters, Entanglement of formation of an arbitrary state of two qubits, Phys. Rev. Lett. {\bf 80}, 2245 (1998).
\bibitem{eberly}T. Yu and J. H. Eberly, Evolution from Entanglement to Decoherence of Bipartite Mixed ``X'' States, Quant. Inf. and Comp. {\bf 7}, 459-468 (2007).

 
 \bibitem{osbornesupp}T. J. Osborne, 
Entanglement for rank-2 mixed states, 
Phys. Rev. A {\bf 72}, 022309 (2005). 


\end{thebibliography}
\end{document}